\documentstyle[preprint,eqsecnum,prb,aps]{revtex}

\begin{document}

\title{Electronic stopping power of aluminum crystal}
\author{I. Campillo, J. M. Pitarke}
\address{Materia
Kondentsatuaren Fisika Saila, Zientzi Fakultatea, Euskal Herriko 
Unibertsitatea,\\ 644 Posta kutxatila, 48080 Bilbo, Basque Country, Spain}
\author{A.G. Eguiluz}
\address{Department of Physics and Astronomy, The University of Tennessee,
Tennessee 37996-1200\\ and Solid State Division, Oak Ridge National Laboratory,
Oak Ridge, Tennessee 37831-6032}

\date\today
\maketitle

\begin{abstract}

Ab initio calculations of the electronic energy loss of ions moving in aluminum
crystal are presented, within linear-response theory, from a
realistic description of the one-electron band-structure and a full treatment
of the dynamical electronic response of valence electrons. For the evaluation
of the density-response function we use the random-phase approximation and,
also, a time-dependent extension of local-density functional theory. We evaluate
both position-dependent and random stopping powers, for a wide range of
projectile velocities. Our results indicate that at low velocities band structure
effects slightly enhance the stopping power. At velocities just above the
threshold velocity for plasmon excitation, the stopping power of the real solid
is found to be smaller than that of jellium electrons, corrections being of
about $10\%$. This reduction can be understood from sum rule arguments.

\end{abstract}


\newpage

\section{Introduction}

The stopping power for charged particles penetrating a solid has been
the topic of considerable theoretical and experimental interest, since the
beginning of this century\cite{Bohr,Bethe,Bloch,Fermi}. The
electronic stopping power due to collisions with valence electrons
has been evaluated for many years on the basis of a jellium
model\cite{Lindhard,Ritchie} of the target, the electronic states being described
by plane waves. However, in a more realistic approach valence electrons move in
a periodic potential, electronic states are described by Bloch wave functions,
and the spectrum of one-electron excitations splits into the so-called energy
bands. The impact of band structure effects on both plasmon dispersion
curves\cite{Quong} and dynamical structure factors\cite{Godby1,Godby2,Fleszar}
has been investigated recently, demonstrating the importance of these
effects even in the case of free-electron-like metals such as aluminum.

Among the most recent attempts to fully introduce the electronic band
structure in the evaluation of the stopping power for low
projectile velocities there is, for alkaline metals, a one-band
calculation\cite{Grande}, as well as a calculation based on a linear combination
of atomic  orbitals\cite{Dorado}. For arbitrary incident velocities the 
stopping power
can be calculated, within linear-response theory, from the knowledge of the
dynamical density-response function of the target\cite{Adler,Wiser,Saslow}.
Approximate semiempirical treatments of this quantity have been made, and
stopping powers of silicon\cite{Kom,Des,Esbensen,Kom2} and
gold\cite{Crawford} for channeled ions have been predicted. More recently, the
low velocity limit has been investigated, on the basis of a static treatment of
the density-response\cite{Tielens}. First-principles treatments of the full
dynamical electronic response of various solids have also been performed, and
preliminary ab initio evaluations of the stopping power of real solids have been
presented\cite{Pitarke,Trickey,Igor}.

In this paper we investigate, within linear-response theory, the
valence electronic energy loss of ions moving through aluminum crystal.
Aluminum is well known to be a typical jellium-like metal with a well-defined
excitation spectrum, and it represents, therefore, an appropriate benchmark
for ab initio evaluations of the electronic stopping power of real solids.
First of all, in section II we describe our full treatment of the wave-vector and
frequency dependent electronic response of valence electrons, based on a
realistic description of the one-electron band structure and first-principles
pseudopotential theory. For the evaluation of the density-response function we
use the random-phase approximation (RPA)\cite{Pines} and, also, a
time-dependent extension of local-density functional theory
(TDLDA)\cite{Soven}. In section III, we derive explicit expressions for both
random and position-dependent stopping powers, from the knowledge of the
imaginary part of the projectile self-energy, which is evaluated in the so-called
GW approximation\cite{Hedin}. In section IV, numerical calculations of the
stopping power of valence electrons in aluminum crystal are presented, for a
wide range of projectile velocities, and we compare our results with the
stopping power of a homogeneous electron gas with a density equal to that of
aluminum. We evaluate, separately, contributions to the energy loss coming from
the excitation of single electron-hole pairs and plasmons, and we interpret our
results on the basis of the f-sum-rule for the dynamical structure
factor. We only consider the electronic response of valence electrons, and the
contribution to the electronic stopping power coming from the excitation of core
electrons is, therefore, not taken into account. In section
V, our conclusions are presented.

\section{DENSITY-RESPONSE FUNCTION}

The linear density-response function $\chi({\bf r},{\bf r}',\omega)$ of an
electron system is defined by the equation
\begin{equation}\label{eq1}
\rho^{ind}({\bf r},\omega)=\int{\rm d}^3{\bf
r}'\chi({\bf r},{\bf r}',\omega)V^{ext}({\bf r}',\omega),
\end{equation}
where $\rho^{ind}({\bf r},\omega)$ is the electron density induced by an 
external potential $V^{ext}({\bf r},\omega)$.

In a {\it self-consistent-field} theory, the induced electron density is derived
from the response function
$\chi^0({\bf r},{\bf r}',\omega)$ for non-interacting electrons moving in an
effective potential
$V_{eff}({\bf r},\omega)$, as follows
\begin{equation}\label{eq2}
\rho^{ind}({\bf r},\omega)=\int
{\rm d}^3{\bf r}'\chi^0({\bf r},{\bf r}',\omega)[V^{ext}({\bf r}',\omega)
+V^{ind}({\bf r}',\omega)],
\end{equation}
where $V^{ind}({\bf r},\omega)$ represents the linear change in 
$V_{eff}({\bf r},\omega)$ 
brought about by
the induced electron density itself. Since we
consider a time-dependent  external field, we are in the more general scenario
of time-dependent  density functional theory (DFT)\cite{Gross}, whose
theorems\cite{Runge} generalize those of the usual DFT\cite{Kohn}.
  
Within time-dependent DFT, $V^{ind}({\bf r},\omega)$ consists of the sum of two
terms (we use atomic units throughout, i. e., $m_e=e=\hbar=1$): The Hartree
contribution,
\begin{equation}\label{eq3}
V_H^{ind}({\bf r},\omega)=\int{\rm d}^3{\bf r}'{1\over |{\bf r} -
{\bf r}'|}\rho^{ind}({\bf r}',\omega),
\end{equation}
which accounts for the average (long-range) effects
of the Coulomb interaction between the target electrons, and the
exchange-correlation (XC) contribution, 
\begin{equation}\label{eq4}
V_{xc}^{ind}({\bf r},\omega)=\int{\rm
d}^3{\bf r}'{{\delta V_{xc}[\rho]}\over
\delta\rho({\bf r},\omega)}\rho^{ind}({\bf r}',\omega) ,
\end{equation}
which accounts for the effects of all many-body short-range correlations not
included in the Hartree approximation. Here, $V_{xc}$ represents the functional
derivative of the XC energy functional $E_{xc}$:
\begin{equation}\label{eq4p}
V_{xc}[\rho]={\delta E_{xc}[\rho]\over
\delta{\rho({\bf r},\omega)}}.
\end{equation}

Introduction of Eqs. (\ref{eq3}) and (\ref{eq4}) into Eq. (\ref{eq2})
and comparison with Eq. (\ref{eq1}) result in the following integral equation for
the self-consistent-field density-response function:
\begin{equation}\label{eq5}
\chi({\bf r},{\bf r}';\omega)=\chi^0({\bf r},{\bf r}';\omega)
+\int{\rm d}{\bf r}_1\int{\rm d}{\bf r}_2\chi^0({\bf r},{\bf r}_1;\omega)
V({\bf r}_1,{\bf r}_2;\omega)
\chi({\bf r}_2,{\bf r}';\omega),
\end{equation}
where
\begin{equation}\label{eq6}
V({\bf r},{\bf r}';\omega)={1\over|{\bf r}-{\bf r}'|}+K_{xc}[\rho]
\end{equation}
and
\begin{equation}\label{eq6p}
K_{xc}[\rho]={{\delta^2
E_{xc}[\rho]}\over\delta\rho({\bf r},\omega)\delta\rho({\bf r}',\omega)}.
\end{equation}

In the time-dependent Hartree or random-phase approximation,
$V_{eff}({\bf r})$ consists only of the average electrostatic interaction between
the electrons, the induced potential is then given by the Hartree term
exclusively, and $K_{xc}({\bf r},{\bf r}')$ in Eq. (\ref{eq6}) is taken to be
zero. In the TDLDA, the kernel $K_{xc}({\bf r},{\bf r}')$ entering Eq.
(\ref{eq6}) is replaced by
\begin{equation}\label{eq7}
K_{xc}({\bf r},{\bf r}')^{LDA}=\delta({\bf r}-{\bf r}')\left [{{\rm d}
V_{xc}(\rho)\over {\rm d}\rho} \right ]_{\rho=\rho_0({\bf r})}, 
\end{equation}
where $V_{xc}(\rho)$ is the derivative of the XC energy of a
homogeneous electron gas of density $\rho$, which we compute with use of the
Perdew and Zunger parametrization\cite{Perdew}. This so-called TDLDA represents
an adiabatic extension (a zero frequency
$K_{xc}$ is used) to finite frequencies of the local density approximation (LDA)
for XC.

For periodic crystals we introduce a Fourier expansion of the 
density-response function,
\begin{equation}\label{eq8}
\chi({\bf r},{\bf r}',\omega)={1\over\Omega}\sum_{\bf q}^{BZ}\sum_{{\bf
G},{\bf G}'}{\rm e}^{{\rm i}({\bf q}+{\bf G})\cdot{\bf r}}{\rm
e}^{-{\rm i}({\bf q}+{\bf G}')\cdot{\bf r}'}\chi_{{\bf G},{\bf G}'}({\bf
q},\omega),
\end{equation}
where $\Omega$ represents the normalization volume, the first sum 
runs over ${\bf q}$
vectors within the first Brillouin zone (BZ), and ${\bf G}$ and ${\bf G}'$ are
reciprocal lattice vectors. Then, introduction of Eq. (\ref{eq8}) into Eq.
(\ref{eq5}) leads to the following matrix equation:
\begin{equation}\label{eq5b}
\chi_{{\bf G},{\bf G}'}({\bf q},\omega)=\chi^0_{{\bf G},{\bf G}'}({\bf
q},\omega)+
\sum_{{\bf G}''}\sum_{{\bf G}'''}\chi^0_{{\bf G},{\bf G}''}({\bf q},\omega)
V_{{\bf G}'',{\bf G}'''}({\bf q},\omega)
\chi_{{\bf G}''',{\bf G}'}({\bf q},\omega),
\end{equation}
where $V_{{\bf G}'',{\bf G}'''}({\bf q},\omega)$ represent Fourier coefficients
of the interaction potential of Eq. (\ref{eq6}). The Fourier
coefficients of the density-response function of non-interacting electrons have
the well-known form\cite{Eguiluz1}
\begin{eqnarray}\label{eq9}
\chi_{{\bf G},{\bf G}'}^0({\bf q},\omega)=&&{1\over \Omega}\sum_{\bf
k}^{BZ}\sum_{n,n'} {f_{{\bf k},n}-f_{{\bf k}+{\bf q},n'}\over E_{{\bf
k},n}-E_{{\bf k}+{\bf q},n'} +\hbar(\omega + {\rm i}\eta)}\cr\cr
&&\times\langle\phi_{{\bf k},n}|e^{-{\rm i}({\bf q}+{\bf G})\cdot{\bf
r}}|\phi_{{\bf k}+{\bf q},n'}\rangle
\langle\phi_{{\bf k}+{\bf q},n'}|e^{{\rm i}({\bf q}+{\bf G}')\cdot{\bf
r}}|\phi_{{\bf k},n}\rangle,
\end{eqnarray}
where the second sum runs over the band structure for each wave vector 
${\bf k}$ in the
first BZ, $f_{{\bf k},n}$ are Fermi factors, and
$\phi_{{\bf k},n}({\bf r},\omega)$ are single-particle Bloch states.

In the RPA, the one-electron Bloch states entering Eq. (\ref{eq9}) are the
self-consistent eigenfunctions of the one-electron Hartree Hamiltonian, and the
coefficients $V_{{\bf G}'',{\bf G}'''}({\bf q},\omega)$ entering Eq. (\ref{eq5b})
are the Fourier coefficients of the bare Coulomb interaction. In the TDLDA, the
one-electron Bloch states entering Eq. (\ref{eq9}) are the self-consistent LDA
solutions of the Kohn-Sham equation of DFT, and the coefficients 
$V_{{\bf G}'',{\bf G}'''}({\bf q},\omega)$ entering Eq. (\ref{eq5b}) are the
Fourier coefficients of the interaction potential of Eq. (\ref{eq6}) with the
XC kernel of Eq. (\ref{eq7}).

For the evaluation of the one-electron Bloch states, we first expand them in a
plane wave basis,
\begin{equation}\label{eq10}
\phi_{{\bf k},n}({\bf r})={1\over\sqrt\Omega}\sum_{\bf G} u_{{\bf k},n}({\bf
G}){\rm e}^{{\rm i}({\bf k}+{\bf G})\cdot{\bf r}},
\end{equation}
with a kinetic energy cutoff of 12 Rydbergs, which corresponds to
keeping approximately 100 ${\bf G}$-vectors in Eq. (\ref{eq10}). Then we
solve for the  coefficients
$u_{{\bf k},n}$ self-consistently, within a full description of the electron-ion
interaction based on the use of a non-local, norm-conserving ionic
pseudopotential\cite{Hamann}. We subsequently evaluate, from Eq. (\ref{eq9}),
the Fourier coefficients $\chi_{{\bf G},{\bf G}'}^0({\bf q},\omega)$, which is
the most demanding part of the response calculation. Finally, we solve the 
matrix equation (\ref{eq5b}) for the Fourier coefficients of the interacting
response function $\chi_{{\bf G},{\bf G}'}({\bf q},\omega)$, which we obtain in 
both RPA and TDLDA.

\section{ELECTRONIC STOPPING POWER}

We consider a particle of charge $Z_1$ moving in an
inhomogeneous medium. The damping rate of the projectile in the
state $\phi_{0}({\bf r})$ with energy $E_0$ is obtained from the knowledge of the
imaginary part of the self-energy $\Sigma({\bf r},{\bf r}';E_{0})$, according
to:
\begin{equation}\label{eq12}
\gamma=-{2}\int{\rm d}{{\bf r}}\int{\rm d}{{\bf
r}'}\phi_{0}^*({\bf r}){\rm Im}\Sigma({\bf r},{\bf r}';E_{0})
\phi_{0}({\bf r}').
\end{equation}

In the so-called GW approximation, the self-energy is given by\cite{Hedin} 
\begin{equation}\label{eq13}
\Sigma({\bf r},{\bf r}';\omega)={{\rm i}\over2\pi}\int_{-\infty}^{\infty}{\rm d}
\omega'G({\bf r},{\bf r}';\omega-\omega')W({\bf r},{\bf r}';\omega'),
\end{equation}
where $G({\bf r},{\bf r}';\omega)$ represents the Green's function for the
projectile, and
$W({\bf r},{\bf r}',\omega)$ is the screened interaction:
\begin{equation}\label{eq13b}
W({\bf r},{\bf r}';\omega)=v({\bf r}-{\bf r}')+Z_1^2\int{\rm d}{{\bf
r}_1}\int{\rm d}{{\bf r}_2} v({\bf r}-{\bf r}_1)\chi({\bf r}_1,{\bf r}_2,\omega)
v({\bf r}_2-{\bf r}'),
\end{equation}
$v({\bf r}-{\bf r}')$ being the bare Coulomb interaction and
$\chi({\bf r},{\bf r}',\omega)$ the density-response function of the medium.
For periodic crystals, we introduce the Fourier representation given by Eq.
(\ref{eq8}) into Eq. (\ref{eq13b}), and we find that
\begin{equation}\label{eq18}
W({\bf r},{\bf r}',\omega)={Z_1^2\over\Omega}\sum_{\bf q}^{BZ}\sum_{{\bf
G},{\bf G}'}{\rm e}^{{\rm i}({\bf q}+{\bf G})\cdot{\bf r}}{\rm
e}^{-{\rm i}({\bf q}+{\bf G}')\cdot{\bf r}'}v_{{\bf G}}({\bf
q})\left[\delta_{{\bf G},{\bf G}'}+
\chi_{{\bf G},{\bf G}'}({\bf q},\omega)\,v_{{\bf G}'}({\bf q})\right],
\end{equation}
where $v_{{\bf G}}({\bf q})=4\pi/|{\bf q}+{\bf G}|^2$ are the Fourier
coefficients of the bare Coulomb interaction.

Replacing the Green's function entering Eq. (\ref{eq13}) by its zeroth order
approximation, we find the following expression for the damping rate:
\begin{equation}\label{eq14}
\gamma=\int_0^\infty{\rm d}\omega P_\omega,
\end{equation}
where
\begin{equation}\label{eq16}
P_\omega=-2\sum_f\int{\rm d}{\bf r}\int{\rm d}{\bf r}'\phi_f^*({\bf
r}')\phi_{0}^*({\bf r}) {\rm Im} W({\bf r},{\bf r}',E_{0}-E_f)\phi_f({\bf
r})\phi_{0}({\bf r}')
\delta[\omega-(E_0-E_f)],
\end{equation}
the sum running over a complete set of final states $\phi_{f}({\bf r})$ of
energy $E_f$. 

The quantity $P_{\omega}$ entering Eq. (\ref{eq14}) represents the probability
for the projectile to transfer energy $\omega$ to the medium. Consequently, the
stopping power of the medium, i. e., the energy-loss per unit path length of the
projectile, can be obtained as follows:
\begin{equation}\label{eq15}  
-{{\rm d}E\over{\rm d}x}={1\over
v}\int_0^\infty{\rm d}\omega\omega P_\omega, 
\end{equation}
where $P_\omega$ is given by Eq. (\ref{eq16}).

\subsection{Random stopping power}

In the case of heavy projectiles moving at a constant velocity ${\bf v}$ with no
definite impact parameter, the initial
and final states can be described by plane waves:
\begin{equation}\label{eq19}
\phi_0({\bf r})={1\over\sqrt\Omega}{\rm e}^{{\rm i}{\bf q}_0\cdot{\bf r}}
\end{equation}
and
\begin{equation}\label{eq20}
\phi_f({\bf r})={1\over\sqrt\Omega}{\rm e}^{{\rm i}({\bf q}_0-{\bf q})\cdot{\bf
r}},
\end{equation}
where ${\bf q}_0$ and ${\bf q}$ represent the initial momentum of the projectile
and the momentum transfer, respectively.

For heavy projectiles recoil can be neglected, i.e.,
\begin{equation}\label{eq21}
E_0-E_f={\bf q}\cdot{\bf v},
\end{equation}
and after introduction of Eqs. (\ref{eq18}), (\ref{eq19}), (\ref{eq20}) and
(\ref{eq21}) into Eq. (\ref{eq16}) we find the following expression for the
so-called random stopping power: 
\begin{equation}\label{eq26r}
\left[-{{\rm d}E\over {\rm d}x}\right]_{\rm random} = -{8\pi\over\Omega
v}Z_1^2\sum_{\bf q}^{\rm BZ}\sum_{\bf G}{({\bf q}+{\bf
G})\cdot  {\bf v}\over |{\bf q}+{\bf G}|^2}{\rm Im} \epsilon_{{\bf G},{\bf
G}}^{-1}\left[{\bf q},({\bf q}+{\bf G})\cdot  {\bf v}\right],
\end{equation}
where $\epsilon_{{\bf G},{\bf G}'}^{-1}({\bf q},\omega)$ is the inverse
dielectric matrix in momentum space:
\begin{equation}\label{eq27}
\epsilon_{{\bf G},{\bf
G}'}^{-1}({\bf q},\omega)=\delta_{{\bf G},{\bf G}'}+{4\pi\over|{\bf
q}+{\bf G}|^2}\chi_{{\bf G},{\bf G}'} ({\bf q},\omega),
\end{equation}
the density-response matrix $\chi_{{\bf G},{\bf G}'}({\bf q},\omega)$ being
given by Eq. (\ref{eq5b}).

The symmetry of the one-particle Bloch states results in the following
identity: 
\begin{equation}\label{eq28}
\epsilon_{{\bf G},{\bf G}'}^{-1}(S{\bf q},\omega)=\epsilon_{S^{-1}{\bf
G},S^{-1}{\bf G}'}({\bf q},\omega), 
\end{equation}
$S$ representing a point group symmetry operation in the periodic crystal. As a
consequence, the stopping power of Eq. (\ref{eq26r}) can be
evaluated from the knowledge of the dielectric matrix for wave vectors ${\bf q}$
lying in the irreducible element of the Brillouin zone (IBZ):
\begin{equation}\label{eq29}
\left[-{{\rm d}E\over {\rm d}x}\right]_{\rm random} = -{8\pi\over\Omega
v}Z_1^2\sum_{\bf q}^{\rm IBZ}\sum_S\sum_{\bf G}{(S{\bf q}+{\bf G})\cdot  {\bf
v}\over |S{\bf q}+{\bf G}|^2} {\rm Im} \epsilon_{S^{-1}{\bf G},S^{-1}{\bf
G}}^{-1}\left[{\bf q},(S{\bf q}+{\bf G})\cdot{\bf v}\right], 
\end{equation}
the second sum in this equation running over the symmetry
operations generating the wave vectors in the star of each
${\bf q}$. 
 
If the diagonal elements of the inverse dielectric matrix entering Eq.
(\ref{eq26r}) are replaced by the inverse dielectric function of a homogeneous
electron gas\cite{Lindhard}, i. e.,
\begin{equation}
\epsilon_{{\bf G},{\bf
G}}^{-1}({\bf q},\omega)\,\to\,\epsilon_{\rm homog}^{-1}(|{\bf q}+{\bf
G}|,\omega),
\end{equation}
Eq. (\ref{eq26r}) exactly coincides with the well-known formula for the
stopping power of a homogeneous electron gas\cite{Echenique}.

\subsection{Position-dependent stopping power}

In the case of heavy projectiles moving with constant velocity ${\bf v}$ on a
definite trajectory at a  given impact parameter
${\bf b}$, the initial and final states can be described in terms of plane 
waves in the
direction of motion and a $\delta$ function in the transverse
direction\cite{Ritchie0}. Then, introduction of these states, $\phi_0({\bf r})$
and
$\phi_f({\bf r})$, and the screened interaction of Eq. (\ref{eq18}) into Eq.
(\ref{eq16}) gives, after neglecting the projectile recoil, the following result
for the position-dependent stopping power:
\begin{equation}\label{eq29p}
\left[-{{\rm d}E\over {\rm d}x}\right]_{\bf b} = -{8\pi\over\Omega
v}Z_1^2\sum_{\bf q}^{\rm IBZ}\sum_S\sum_{\bf G}\sum_{\bf K}{'}e^{{\rm i}{\bf
K}\cdot{\bf b}}{(S{\bf q}+{\bf G})\cdot  {\bf v}\over |S{\bf q}+{\bf
G}+{\bf K}|^2} {\rm Im} \epsilon_{S^{-1}{\bf G},S^{-1}({{\bf G}+{\bf
K}})}^{-1}\left[{\bf q},(S{\bf q}+{\bf G})\cdot  {\bf v}\right], 
\end{equation}
the sum $\displaystyle\sum_{\bf K}{'}$ being
restricted to those reciprocal lattice vectors that are perpendicular to the
velocity of the projectile, i. e., ${\bf K}\cdot{\bf v}=0$. $\epsilon_{{\bf
G},{\bf G}'}^{-1}({\bf q},\omega)$ is the inverse dielectric matrix of Eq.
(\ref{eq27}) and
$S$ represents, as in Eq. (\ref{eq29}), a symmetry operation of the point group
of the periodic crystal.
 
The most important contribution to the position-dependent stopping power of
Eq. (\ref{eq29p}) is provided by the term ${\bf K}=0$, the magnitude of the other
terms depending on the direction of the velocity. For those directions for which
the condition of ${\bf K}\cdot{\bf v}=0$ is never satisfied we have the random
stopping power of Eq. (\ref{eq29}), and for a few highly symmetric or {\it
channeling} directions non-negligible corrections to the random result are
found, thus exhibiting the anisotropy of the position-dependent stopping power.
We also note that (a) the average over impact parameters of the
position-dependent stopping power of Eq. (\ref{eq29p}) along any given channel
has the same value as the random stopping power of Eq. (\ref{eq29}), and (b) as
long as the diagonal elements of the so-called dynamical structure factor,
$-2\,{\rm Im}\chi_{{\bf G},{\bf G}}({\bf q},\omega)$, were isotropic, there
would be no dependence of the random stopping power on the direction of ${\bf
v}$.

\section{RESULTS and DISCUSSION}

The input of our calculation of both random and position-dependent
stopping powers of Eqs. (3.14) and (3.16) is the interacting response
matrix $\chi_{{\bf G},{\bf G}'}({\bf q},\omega)$ of Eq. (2.11), which we have
solved for Al crystal. The number of bands required in the evaluation of the
polarizability $\chi_{{\bf G},{\bf G}'}^0({\bf q},\omega)$ of Eq. (2.12)
depends on the value of the frequency $\omega$. As the maximum energy $\omega$
transferred by the moving projectile to the target is proportional to the
velocity, only a few bands are required in the case of slow projectiles.
Indeed, it has been found\cite{Pitarke} that at velocities below $v=0.1{\rm
a.u.}$ just two bands are required to account for the energy-loss. At larger
velocities excitations to higher bands become possible, and we find that
convergence is achieved, for $v<2{\rm a.u.}$, if one considers on the order of 30
bands. The results presented below have been found to be well-converged for all
velocities under study, and they all have been performed with the use of 60
bands.

The sampling of the BZ required for the evaluation of both the
polarizability of Eq. (2.12) and the stopping powers of Eqs. (3.14) and (3.16)
has been performed on $10\times 10\times 10$ Monkhorst-Pack meshes\cite{MP} ($47$
points in the IBZ) with an energy smearing of the Fermi level of $0.25{\rm eV}$,
and this sampling has been done in conjunction with a choice of a finite value of
the damping factor entering Eq. (2.12) of $\eta=1.5{\rm eV}$. At velocities of
the projectile below the velocity for which plasmon excitation becomes possible,
convergence is already achieved with the use of $4\times 4\times 4$
Monkhorst-Pack meshes ($8$ points in the IBZ).

\subsection{Random stopping power}

The main ingredient in the calculation of the random stopping power of Eq.
(3.14) is the so-called dynamical structure factor $S_{\bf G}({\bf q},\omega)$.
It is proportional to the diagonal elements of the imaginary part of the
density-response matrix:
\begin{eqnarray}\label{s}
S({\bf q}+{\bf G},\omega)&=&-2{\rm Im}\chi_{{\bf G},{\bf G}}({\bf
q},\omega)\cr\cr &=&-{|{\bf q}+{\bf G}|^2\over 2\pi}{\rm
Im}\epsilon^{-1}_{{\bf G},{\bf G}}({\bf q},\omega),
\end{eqnarray}
where ${\bf q}$ is taken to be in the first BZ. Thus, couplings of the wave
vector
${\bf q}+{\bf G}$ to wave vectors ${\bf q}+{\bf G}'$ with ${\bf G}\neq{\bf G}'$,
which appear as a consequence of the existence of electron density variations in
real solids, only contribute to the random stopping power through the dependence
of the diagonal elements of the interacting response matrix
$\chi_{{\bf G},{\bf G}}({\bf q},\omega)$ on the off-diagonal elements of the
polarizability
$\chi_{{\bf G},{\bf G}'}^0({\bf q},\omega)$. We have found that in the case of Al
crystal, which does not present strong electron density gradients nor special
electron density directions (bondings), contributions from these so-called
crystalline local field effects are within $0.5\%$ of the total random stopping
power. Thus, as for the evaluation of the random stopping power of Al crystal,
the off-diagonal elements of $\chi_{{\bf G},{\bf G}'}^0({\bf q},\omega)$ can be
neglected. 

In the general case of an inhomogeneous system one finds, for the imaginary part
of the inverse dielectric function, the following form of the f-sum-rule:
\begin{equation}\label{sum}
\int_0^{\infty}{\rm d}\omega\omega{\rm Im}\epsilon_{{\bf G},{\bf G}'}^{-1}({\bf
q},\omega)= -{2\pi^2}n_{{\bf G}-{\bf G}'},
\end{equation}
where $n_{{\bf G}}$ represent the Fourier components of the density.
Note that $n_{{\bf G}=0}$, which equals the average density of the system, does
not depend on the details of the band structure. Thus, it may be argued that band
structure effects, which have an impact on both plasmon dispersion\cite{Quong}
and the dynamical structure factor\cite{Fleszar}, may give no corrections
to integrated quantities as the random stopping power. However, because in the
sum over the wave vector ${\bf q}$ in Eq. (3.14) the frequency $\omega$ in the
argument of the structure factor takes values from 0 to only $|{\bf q}+{\bf
G}|\,v$, the stopping power of band electrons and that of electrons in an
effective jellium with the same average electron density will be, in general,
different.

In Fig. 1 we show, as a function of the velocity of the projectile,
our full calculation of the random stopping power (see Eq. (3.14)) of valence
electrons in Al crystal for protons ($Z_1=1$), together with the corresponding
result for a homogeneous electron gas with an electron density parameter equal
to that of aluminum ($r_s=2.07$)\cite{rs}. In the calculation of the stopping
power of both band and jellium electrons, the third sum in Eq. (3.14) is
extended over
$15{\bf G}$ vectors of the reciprocal lattice, the magnitude of the maximum
momentum transfer ${\bf q}+{\bf G}$ being $2.9q_F$ ($q_F$ is the Fermi
momentum)\cite{note1,note2}. Solid and open circles represent our calculated
random stopping powers of band electrons, as calculated within the RPA and the
TDLDA, respectively. Solid and dashed lines represent the corresponding stopping
powers of jellium electrons. All these results have been found, for Al crystal,
to be insensitive to the choice of the projectile velocity direction.

In the case of projectiles moving at low velocities, below the threshold velocity
for plasmon excitation, the sum over the frequency
$\omega=({\bf q}+{\bf G})\cdot{\bf v}$ in Eq. (3.14) can never be replaced by an
integration from $0$ to $\infty$ as in Eq. (\ref{sum}), and the stopping power
will, in principle, depend on the band structure of the crystal. The existence of
interband transitions, not present within a jellium model of the crystal,
result in a dynamical structure factor which is, at low frequencies (where it
increases linearly with frequency), slightly enhanced with respect to the
corresponding jellium result\cite{Fleszar}. As a result, the stopping power
of the real target is, for projectile velocities smaller than the Fermi
velocity and within both RPA and TDLDA, a linear function of the velocity and
about $7\%$ higher than the stopping power of jellium electrons.

The threshold velocity for which plasmon excitation becomes possible is
$v_t\approx 1.3{\rm a.u.}$ in the case of electrons in jellium with $r_s=2.07$.
In the case of Al crystal this threshold velocity is slightly smaller ($v\approx
1.2{\rm a.u.}$), as can be easily concluded from an inspection of the plasmon
dispersion curve for Al crystal of Ref.\onlinecite{Quong}. For projectile
velocities larger than
$v_t$ there are two mechanisms of valence electron excitation: (a) excitation of
single electron-hole pairs and (b) excitation of plasmons. Figure 2
exhibits, as a function of the projectile velocity, separate contributions to
the stopping power of both band (solid circles) and jellium (solid lines)
electrons coming from electron-hole pair excitation (Fig. 2a) and plasmon
excitation (Fig. 2b). Although for wave vectors that are smaller than $q_c$ (the
critical wave vector where the plasmon dispersion enters the electron-hole
pair excitation spectrum) both mechanisms of valence electron excitation
contribute to the energy-loss, contributions from losses to electron-hole
pair excitations are negligible for $q<q_c$, and Eq. (4.2) leads us
to the conclusion that contributions from losses to plasmon excitation
are independent of the detailed band structure of the crystal. That this is the
case is obvious from Fig. 2b. As for the contribution to the energy-loss coming
from the excitation of electron-hole pairs, Figs. 1 and 2a show that band
structure corrections lower the stopping power of electrons in jellium by about
$10\%$ at and just above the plasmon threshold velocity; this is a consequence of
the dynamical structure factor for the real crystal being, within the
electron-hole pairs continuum at $q>q_c$ and
$\omega<(q+G)\,v$, smaller than in the
case of a homogeneous electron gas. Nevertheless, contributions to the stopping
power coming from the excitation of plasmons are still smaller than
contributions from losses to electron-hole pairs, and there is, at high
velocities, exact equipartition of the energy-loss, as in the case of a
homogeneous electron gas\cite{LindhardW}. Also, at high velocities the sum over
the frequency
$\omega=({\bf q}+{\bf G})\cdot{\bf v}$ in Eq. (3.14) can be replaced by an
integration from $0$ to $\infty$ and the stopping power of both jellium and band
electrons is expected to be the same, as illustrated in Figs. 1 and 2a.  
 
We note that the TDLDA results plotted in Fig. 1 for the stopping power of
jellium electrons, which describe short-range correlation effects within the LDA,
approximately reproduce more detailed calculations with either static or dynamic
local-field corrections included\cite{Ping}. The stopping power of both jellium
and band electrons is enhanced by about
$20\%$, at low velocities (exclusive electron-hole pairs domain), with respect to
the RPA result, as a consequence of short-range correlation effects provoking a
reduction in the screening within the target. The
sum rule of Eq. (\ref{sum}) equally applies for both RPA and TDLDA dynamical
structure factors, and this prevents, therefore, plasmon contributions to the
stopping power from being sensitive to the details of the wave-vector and
frequency dependence of the response. At large velocities, well above the
stopping maximum, all calculations presented in Fig. 1 converge to the
well-known Bethe formula\cite{Echenique}, in which the mean excitation potential
is replaced by the plasma energy\cite{note3}.

\subsection{Position-dependent stopping power}

We have carried out, from Eq. (3.16), calculations of the electronic
energy-loss versus impact parameter, and we have obtained, within the RPA, the
results presented in Fig. 3. In this figure we represent by solid curves the
position-dependent stopping power of Al crystal for protons ($Z_1=1$) moving
with $v=0.2{\rm a.u.}$ along the (100) (Fig. 3a) and (111) (Fig. 3b)
directions. In Fig. 3a the impact parameter has been taken to be ${\bf
b}=(0,b,0)a_c$ ($a_c$ is the lattice constant), with $b$ from $0$ to $1$, and
in Fig. 3b it has been taken to be ${\bf b}=(-1\,1\,0)a_c$. Contour density plots
of the square lattice containing both the projectile trajectory and the ${\bf
b}$ vector are displayed in Figs. 4a and 4b for the (100) and (111)
directions, respectively. The averaged electronic densities along the projectile
paths depicted in Fig. 4 are plotted in Fig. 5, as a function of the
impact parameter, showing for this fcc crystal the expected periodicity with
$b=1/2$, also present in the position-dependent stopping power of Fig 3.

First of all, we note that the existence of small electron density
variations in real aluminum results, through the off-diagonal elements of the
interacting density-response function $\chi_{{\bf G},{\bf G}'}({\bf q},\omega)$,
in non-negligible differences between position-dependent and random stopping
powers. We have obtained differences up to $20\%$ for projectiles incident in the
(100) direction (see Fig. 3a), and up to $10\%$ for projectiles moving in the
(111) direction (see Fig. 3b). The maxima in the stopping power for trajectories
along the interstitial regions ($b=1/4$) and the minima near the cores (b=0) are
associated with corresponding maxima and minima in the integrated electronic
densities of Fig. 5.

A comparison with a local density approximation (LDA)\cite{note4} is also shown
in Fig. 3 (dashed curves). In this approach, the position-dependent stopping
power is obtained as the stopping power of a homogeneous electron gas with an
electron density equal to the average electron density along the projectile path.
Although the impact parameter dependence, as obtained in this approach, is
qualitatively the same as found within our full band structure calculation,
position-dependent corrections to the random stopping power appear to be
largely underestimated within the LDA. This is obviously a consequence of
the stopping power depending not only on the local density, but also on the
electron density gradients further away from the projectile trajectory.
We have also calculated position-dependent stopping powers of Al
crystal for various non-relativistic velocities of the incident projectile, and
we have found that corrections to the random stopping power, as obtained within
the LDA, are too small for all velocities. These corrections are found, within
our full treatment of the band structure, to slightly decrease with the
velocity of the projectile. In the case projectiles moving near the cores
($b=0$) along the (100) direction, we have obtained differences between
position-dependent and random stopping powers of approximately $20\%-12\%$ in the
velocity range from $v=0.2{\rm a.u.}$ to $v=1{\rm a.u.}$. For projectiles moving
near the cores ($b=0$) along the (111) direction, these differences have been
found to be, in the same velocity range, of approximately $10\%-8\%$.

\section{Conclusions}

We have presented full band structure calculations of both random and
position-dependent stopping powers of valence electrons in Al crystal, as
obtained within linear-response theory and with the use of the RPA and the
TDLDA. The random stopping power has been evaluated for a wide range of
projectile velocities. Our results indicate that at low velocities of the
projectile band structure effects result in the stopping power of real aluminum
being about $7\%$ higher than that of a homogeneous electron gas with a density
equal to the average electron density of the real crystal ($r_s=2.07$). At
velocities just above the threshold velocity for plasmon excitation, the
stopping power of valence electrons in Al crystal is found to be smaller than
that of jellium electrons, corrections being of about $10\%$. In the
high-velocity limit, the stopping power of both jellium and band electrons is
found to be the same and to coincide with the well-known Bethe formula.

The various contributions to the random stopping power of Al crystal have
been calculated separately, showing that band structure effects on the
contribution to the stopping power coming from plasmon excitation are
negligible. This result has been shown to be a consequence of the dynamical
structure factor of both jellium and real aluminum fulfilling the same sum
rule, as long as the average electron density is the same. Also, there is, in
the high-velocity limit, exact equipartition for the energy-loss, as in the case
of a homogeneous electron gas.

The position-dependent stopping power has been evaluated for projectiles with
velocities up to the Fermi velocity incident in various high-symmetry
directions. We have found differences between position-dependent and random
stopping powers up to $10\%$ for projectiles incident in the (100) direction
and up to $20\%$ for projectiles moving in the (111) direction. The magnitude
of these position-dependent relative corrections to the random stopping power
has been found to be, within the LDA, largely underestimated, specially at
high velocities of the projectile.

\acknowledgments

We thank Alberto Garc\'\i a for his help in the calculation of the
one-electron Bloch states. I.C. and J.M.P. acknowledge partial support by the
Basque Unibertsitate eta Ikerketa Saila and the Spanish Ministerio de
Educaci\'on y Cultura. A.G. Eguiluz acknowledges support from National Science
Foundation Grant No. DMR-9634502 and from the National Energy Research
Supercomputer Center. ORNL is managed by Lockheed Martin Energy Research Corp.
for the U.S. DOE under contract No. DE-AC05-96OR22464.

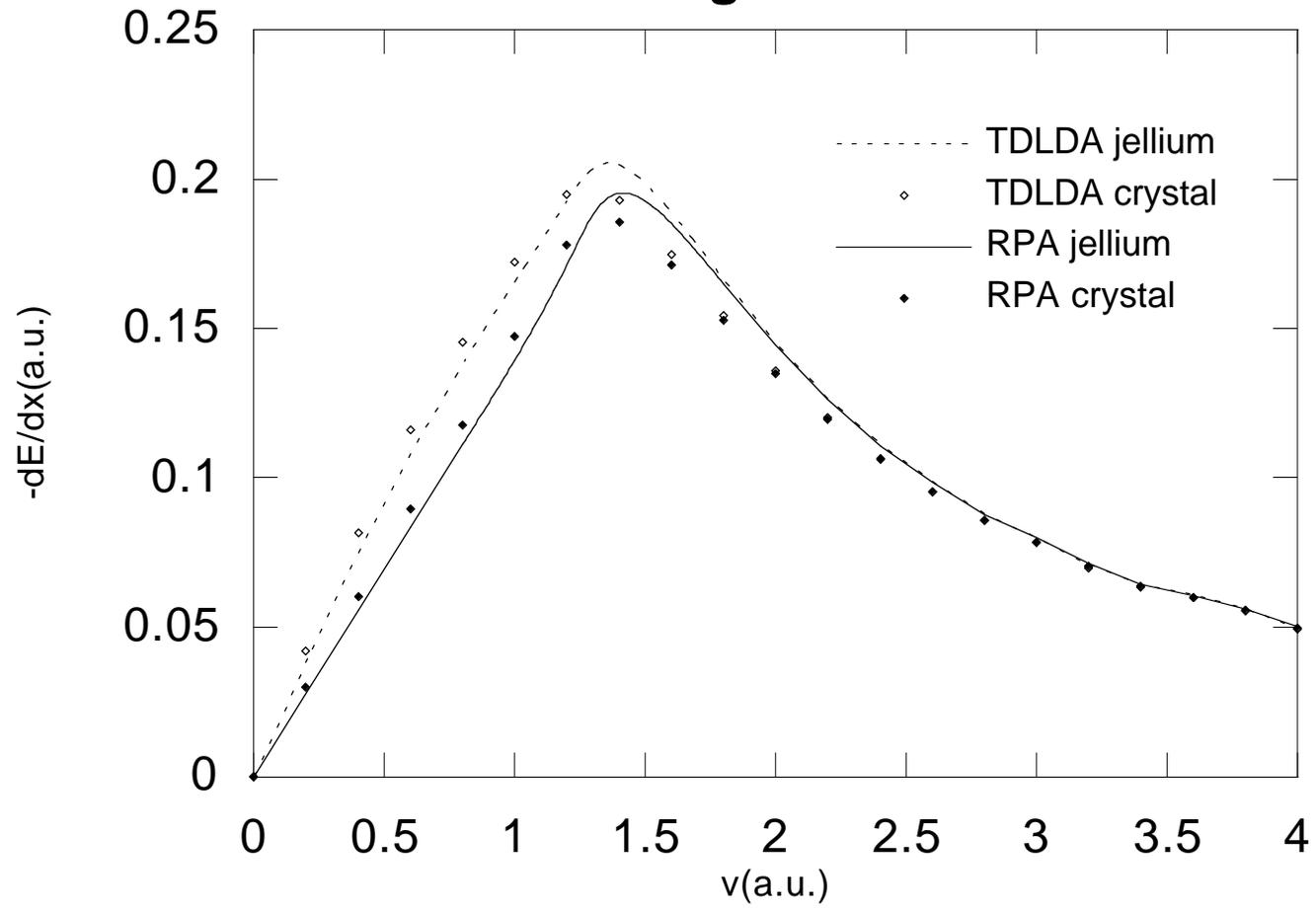
\begin{figure}
\caption{Full band structure calculation, from Eq. (3.14), of the random
stopping power of valence electrons in Al crystal for protons ($Z_1=1$), as a
function of the projectile velocity. Solid and open circles represent the
results obtained in the RPA and the TDLDA, respectively. RPA and TDLDA stopping
powers of electrons in an homogeneous electron gas with $r_s=2.07$ are
represented by solid and dashed lines, respectively.}
\end{figure}

\begin{figure}
\caption{Contributions to the random stopping power of valence electrons in Al
crystal coming from the excitation of (a) electron-hole pairs and (b) plasmons,
versus the projectile velocity, as obtained from Eq. (3.14) within
the RPA. Full band structure calculations are represented by solid circles. The
corresponding jellium results are represented by solid lines.}
\end{figure}

\begin{figure}
\caption{Position-dependent stopping power of Al crystal for protons moving
with $v=0.2$, as obtained within the RPA. The projectile is
incident in the (l\,n\,m) direction and the stopping power is plotted as a
function of the magnitude of the impact vector along the
($\alpha\,\beta\,\gamma$) direction, with (a) (l\,n\,m)=(100) and
($\alpha\,\beta\,\gamma$)=(010); (b) (l\,n\,m)=(111) and
($\alpha\,\beta\,\gamma$)=(-1\,1\,0). Solid lines represent full band structure
calculations, as obtained from Eq. (3.16). Dashed curves represent LDA
calculations, obtained as explained in the text, and the horizontal dashed line
represents the stopping power of a homogeneous electron gas with
$r_s=2.07$. All results have been normalized to the value of the
random stopping power of Al crystal for the same velocity.}
\end{figure}

\begin{figure}
\caption{Contour valence-density plots of the square lattice containing the
projectile trajectory (l\,n\,m) and the impact vector ($\alpha\,\beta\,\gamma$),
with: (a) (l\,n\,m)=(100) and ($\alpha\,\beta\,\gamma$)=(010); (b)
(l\,n\,m)=(111) and ($\alpha\,\beta\,\gamma$)=(-1\,1\,0). Projectile
trajectories with $b=1/4$ are also represented.}
\end{figure}

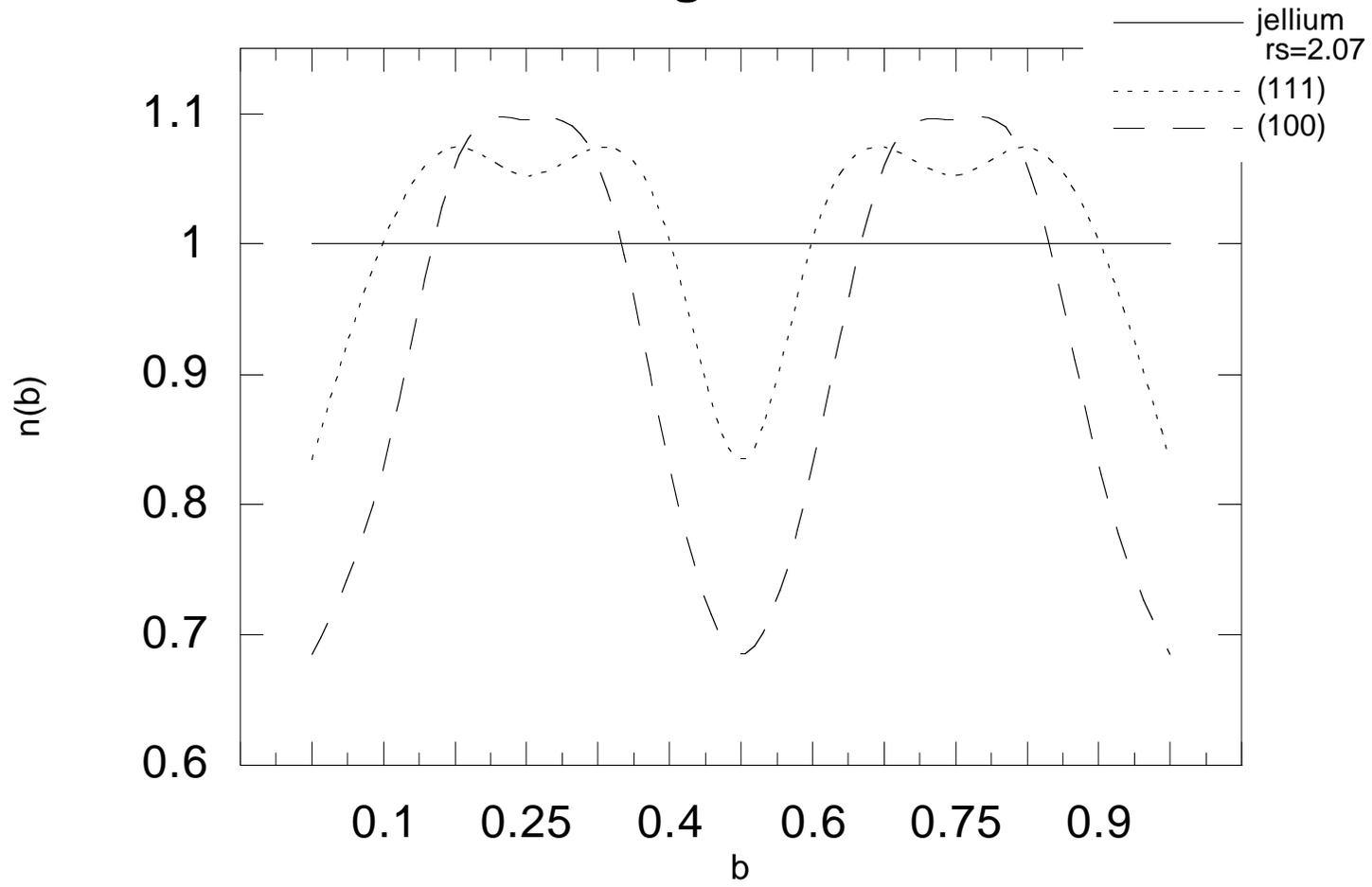
\begin{figure}
\caption{Averaged valence-densities along the projectile paths depicted in
Fig. 4, as a function of the impact parameter $b$. The solid line corresponds
to a projectile trajectory along the (100) direction, as in Figs. 3a and 4a.
The dashed line corresponds to a projectile trajectory along the (111)
direction, as in Figs. 3b and 4b. All results have been normalized to the value
of the average valence-density of Al, $n_0$.}
\end{figure}


\begin{references}
\bibitem[39]{note2} Contributions to the stopping power of a homogeneous
electron gas are well-known to come, within the RPA, from all wave vectors which
are smaller that $2(q_F+v)$, and our calculated results of Fig. 1 are, therefore,
expected to underestimate the total stopping power of both jellium and band
electrons for all velocities $v>0.5{\rm a.u.}$. However, we have found that
differences, at large wave vectors, between the integrated dynamical structure
factor for jellium and band electrons are very small, and the results of Fig. 1
provide, therefore, an accurate measure of the effect of the band structure on
the random stopping power of valence electrons.
\bibitem[42]{note3} The Bethe formula is, in the high velocity limit,
nearly insensitive to differences among the plasma frequencies obtained in the
various approximations to the response of the target.
\bibitem[43]{note4} The use of this so-called local density approximation (LDA)
should not be confused with the use, within DFT, of the LDA for the evaluation of
$V_{xc}$ and the kernel $K_{xc}$ of Eqs. (2.5) and (2.8).
\end{references}
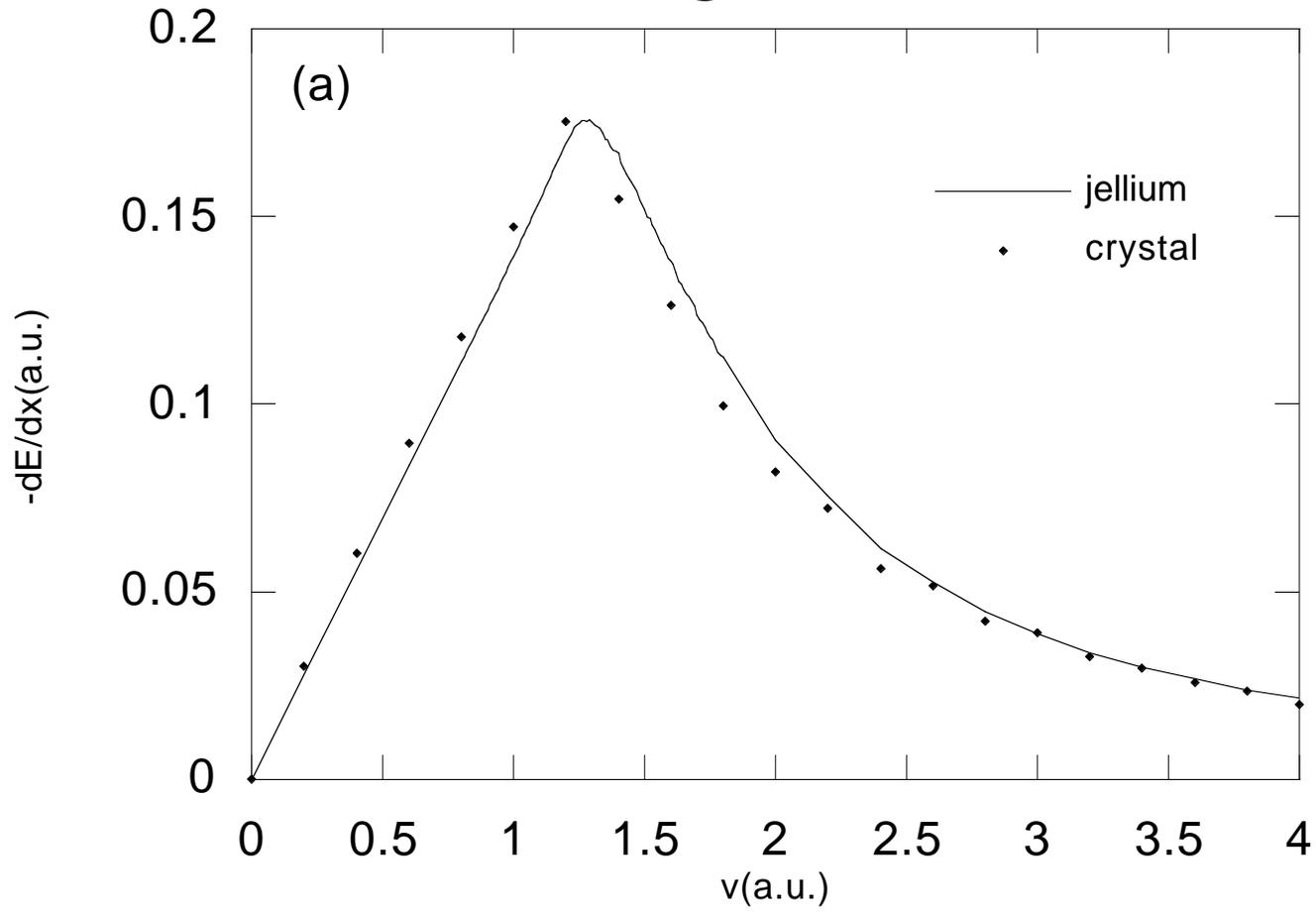
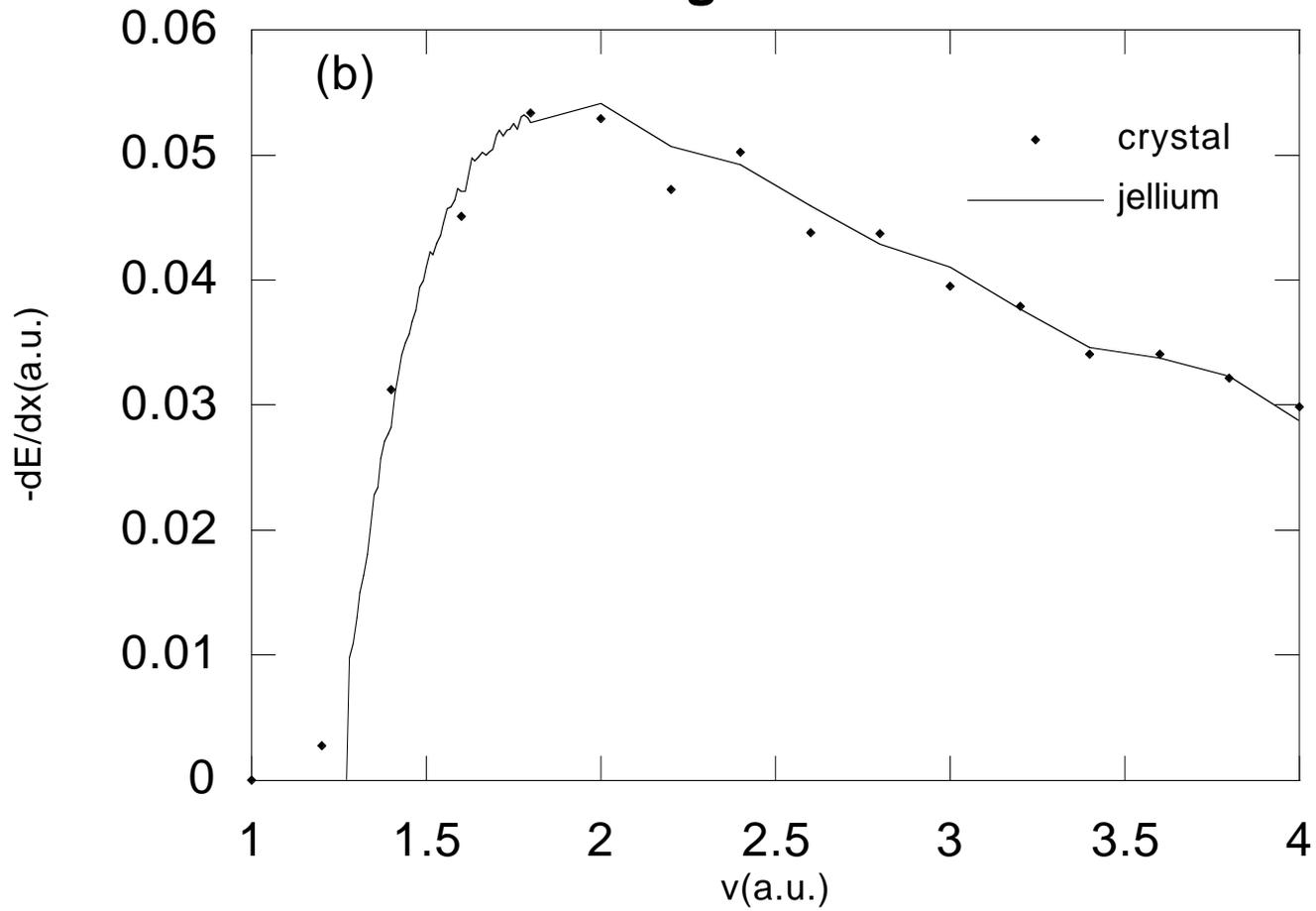
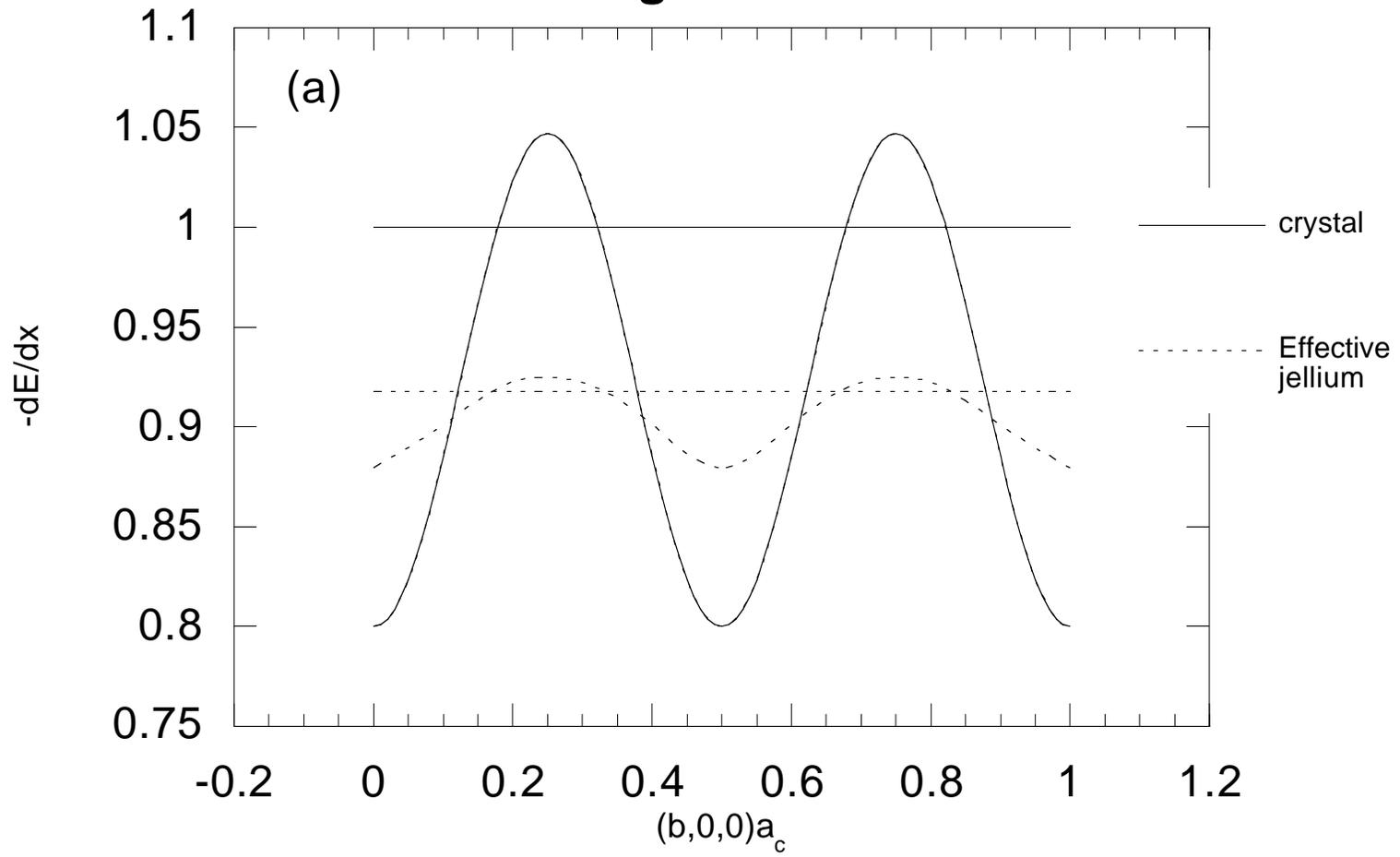
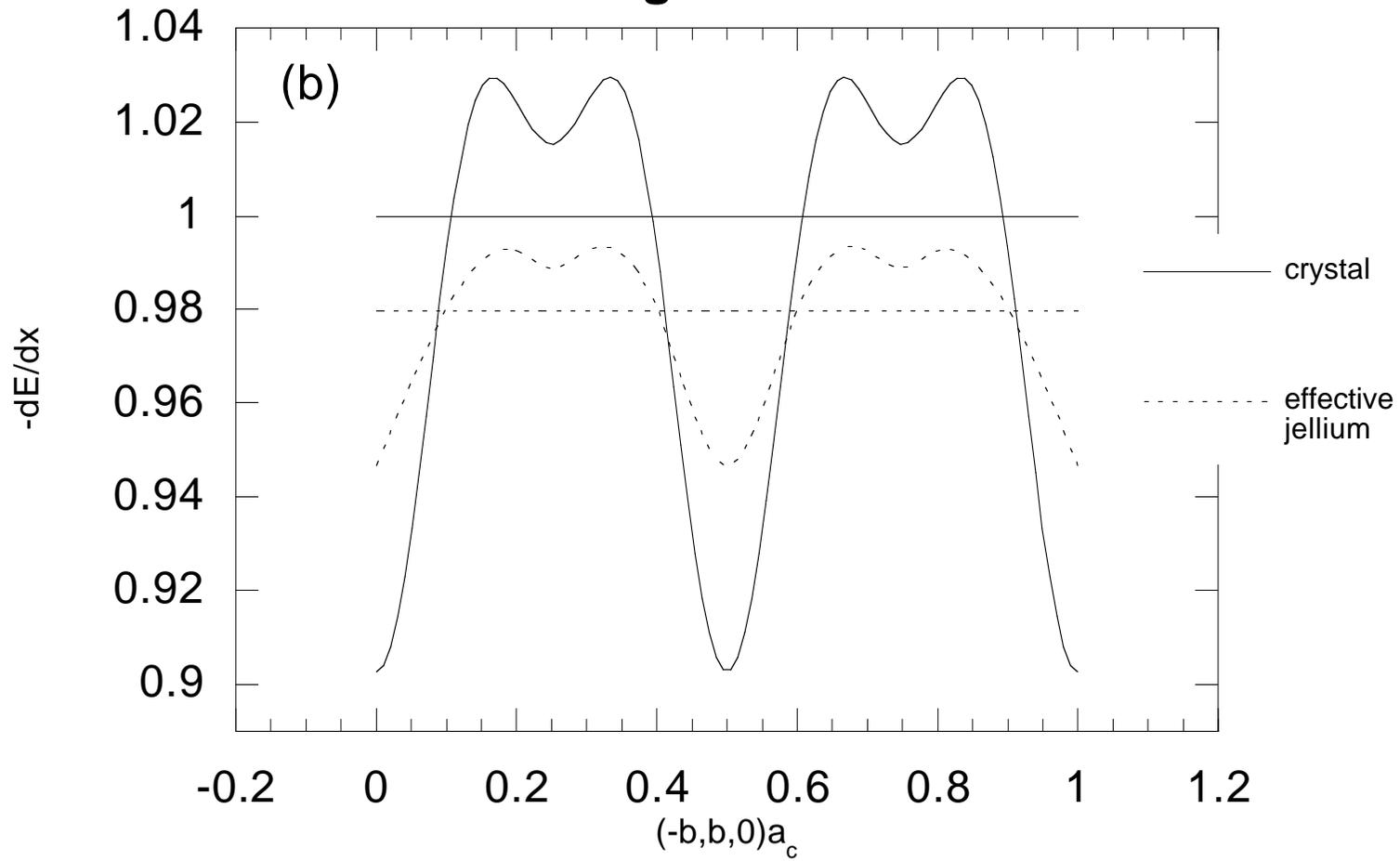
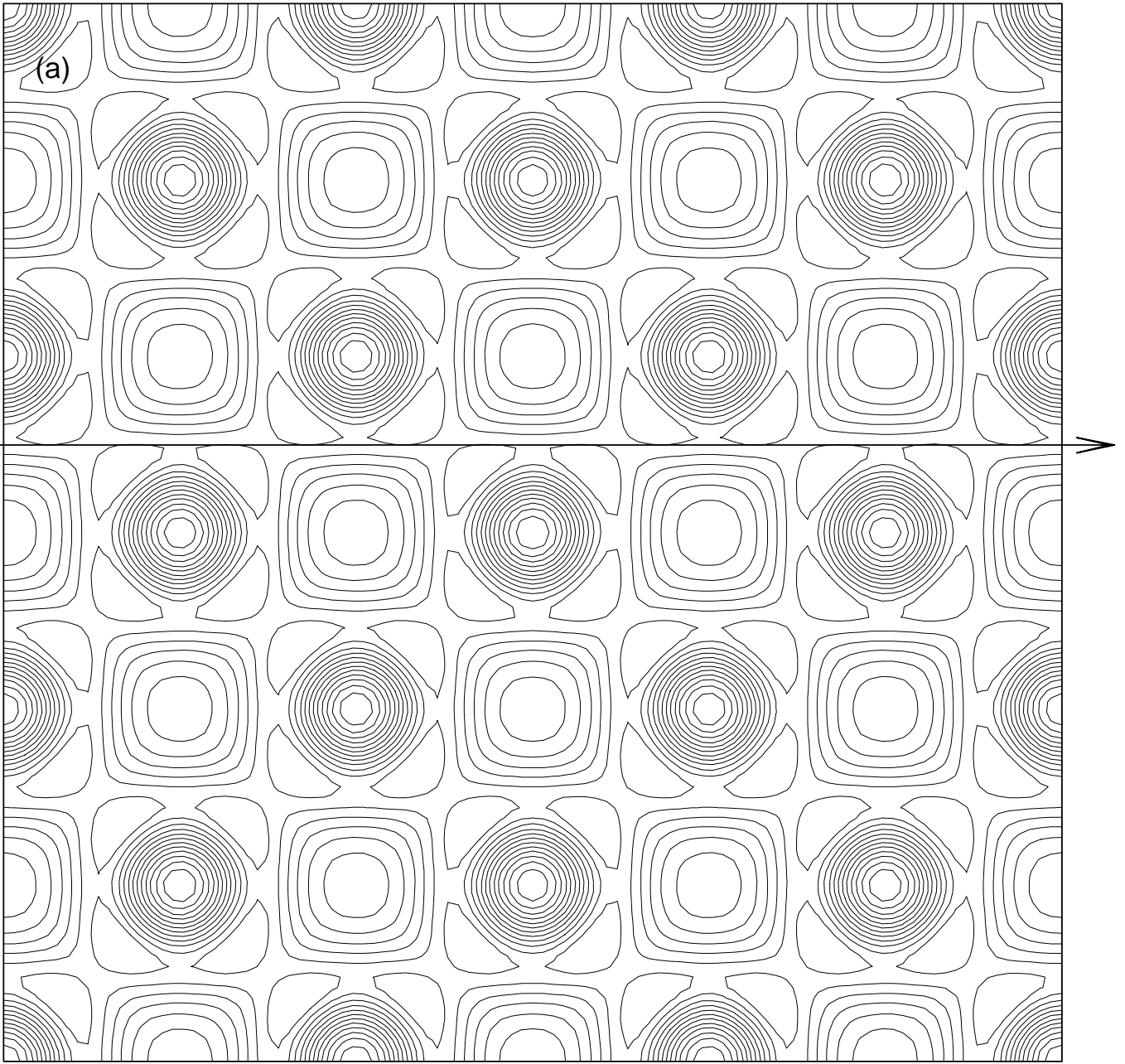
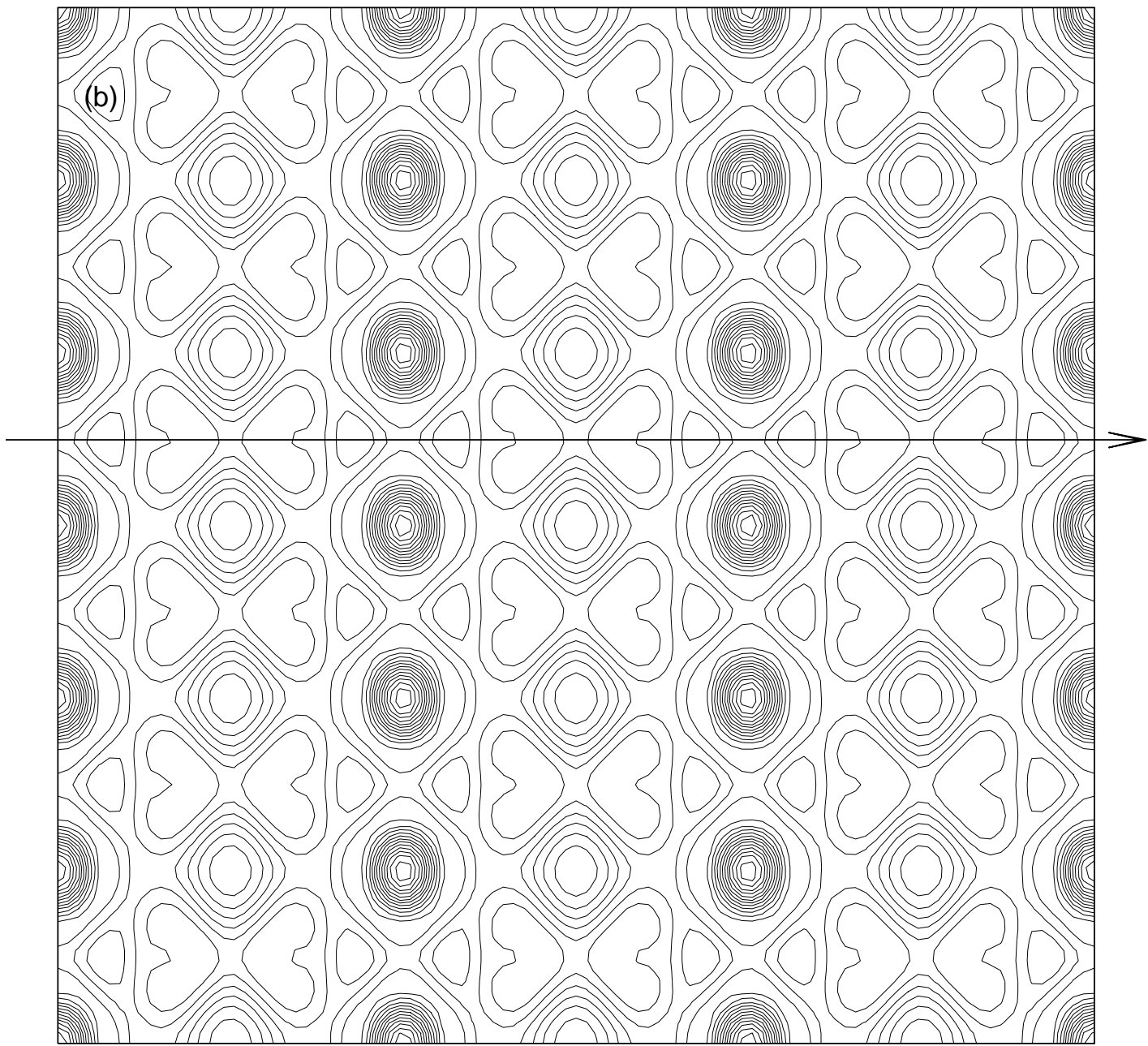
\end{document}